\documentclass[prl,twocolumn,epsf,psfig]{revtex4}
\usepackage{graphicx}
\DeclareGraphicsExtensions{.pdf,.png,.gif,.jpg,.eps}
\usepackage{float}
\usepackage{subfig}

\begin{document}

\title{Spin diffusion of lattice fermions in one dimension}

\author{Andrew P. Snyder and Theja N. De Silva}
\affiliation{Department of Physics, Applied Physics and Astronomy,
The State University of New York at Binghamton, Binghamton, New York
13902, USA.}
\begin{abstract}
We study long-time spin diffusion of harmonically trapped lattice fermions in one dimension. Combining thermodynamic Bethe ansatz approach and local density approximation, we calculate spin current and spin diffusion coefficient driven by the population imbalance. We find spin current is driven by susceptibility effects rather than typical diffusion where magnetization would transport from regions of high magnetization to low. As expected, spin transport is zero through insulating regions and are only present in the metallic regions. In the weak coupling limit, the local spin diffusion coefficient shows maxima at all the insulating regions. Further, we estimate damping rate of diffusion modes in the weak coupling limit within the lower metallic portion of the cloud.
\end{abstract}

\maketitle

\section{I. Introduction}

The transport of fermions is central to many areas in physics. Fermion transport can be seen in various systems, including electronic matter, stars during collapses and explosions, and quark-gluon plasma. In particular, the transport of spin is highly relevant to potential technological applications known as spintronics. The subject of spintronics is to design devices to control spin evolution and dynamics for information storage, transfer, and manipulation \cite{spt}. Thanks to the current state of cold gas experiments, cold gas systems provide a unique opportunity to study these transport properties in a controlled fashion.

The ability of engineering optical lattices in arbitrary geometries in different dimensionalities opens up exciting avenues to visualize and gain deeper understanding of fundamental many body physics. Recent progress in experimental techniques of ultra-cold atoms, such as the technique of detecting a single-site~\cite{insm, insmA}, allows one to probe not only the static properties, but also the dynamics of the atoms in optical lattices.

On shorter timescales, a systems of ultracold atoms can be used as an effective simulator for thermal equilibrium physics in closed many-body quantum systems. In contrast, on longer time scales, the presence of underlying inhomogeneous trapping potential provides an ideal platform for studying numerous non-equilibrium phenomena. In cold gas experiments in the presence of population imbalance, inhomogeneous trapping potential creates a density imbalance throughout the lattice. This density imbalance mimics a magnetization and drives the spin current in the cold atom setups. In solid state devices spin current is produced by injecting spins into the device and can thus be probed by magnetic microscopy or neutron scattering. After the recent proposal on the ultra-cold atom analog of electronic semiconductor devices known as "atomtronics"~\cite{spt1, spt2}, the study of transport phenomena gained a tremendous momentum~\cite{spte1, spte2, spte3, spte4, spte5, spte6, spte7, spte8, spte9, spte10}. In the absence of optical lattice, spin transport in strongly interacting Fermi gases has been experimentally investigated in two laboratories at MIT and Rice University~\cite{spte3, spte6}. In general, the non-equilibrium dynamics of cold atoms in optical lattices are studied after an adiabatic or sudden change of the atom or lattice parameters \cite{none1, none2, none3, none4, none5, none6, none7, none8, none9}. While the dynamics for the adiabatic process are expected to follow the change of the system Hamiltonian, the sudden quench leads to the non-equilibrium physics that allows for the understanding of relaxation dynamics in the presence of many-body interactions. Although non-equilibrium dynamics due to the adiabatic or sudden changes have been widely studied, longer time scale dynamics due to the inhomogeneous trapping potential have been largely unexplored.

The spin of the atoms referred to in this paper are one of two hyperfine states available to the atoms. Therefore, we treat the spin as a scalar quantity which points either parallel (up) or antiparallel (down) to some arbitrary direction of quantization. The magnetization $M$ defined below is proportional to the difference in up and down atom densities. As the overall magnetization in the system is produced by the fixed population imbalance and the interactions are spin-independent, the total magnetization is a conserved quantity. The local magnetization cannot disappear locally, but can only relax slowly over the entire system. The purpose of this paper is to investigate how atoms relax by the magnetization being physically transported within the trap. Indeed, this transport process occurs only in the presence of inhomogeneous potential via a diffusion motion in a larger time scale provided by the underlying harmonic potential.

In the present paper, we study the spin diffusion of one dimensional lattice fermions in the presence of an external harmonic potential. We consider longer timescale (defined later) dynamics of population imbalanced fermions and study the spin diffusion current and local spin diffusion coefficient, and how these values effect the damping rate of the spin diffusion modes. We find evidence of spin current in the metallic regions of the lattice, while there being no spin current in insulating regions. Further, in the weak coupling limit, we calculate local spin diffusion coefficient for all regions of the lattice. The damping rate of spin diffusion modes are estimated using the continuity equation.

The structure of the paper is as follows. In section II, we introduce our model and formalism. In section III, we calculate the ratio of spin current to local diffusion coefficient for characteristic values of large and small interaction, and show the interplay between magnetization and susceptibility effects to spin current. Further, the local spin diffusion coefficient is found over the lattice for the weak coupling limit. Finally, the damping rate of diffusion modes is estimated within a metallic portion of the cloud. We display our results with figures in the appropriate dimensionless quantities. In section IV, we summarize the results and connect them to experiments.

\section{II. Formalism}

We consider a cloud of atoms consisting of two hyperfine states denoted by spin $\uparrow$ and $\downarrow$. We assume that the population imbalance $P = (N_\uparrow-N\downarrow)/N$ is finite, where $N = N_\uparrow+N_\downarrow$ is the total number of atoms in two hyperfine states of the same atom. These atoms are subjected to a combined optical lattice and harmonic trapping potential. The Hamiltonian of the system can be represented by the one-dimensional Hubbard model.

\begin{eqnarray}
H = - t\sum_{<ij>,\sigma} c^\dagger_{i\sigma}c_{j\sigma}+U\sum_in_{i\uparrow}n_{i\downarrow} \\ \nonumber
- \mu \sum_{i\sigma} c^\dagger_{i\sigma}c_{i\sigma}-h\sum_{i\sigma} \sigma c^\dagger_{i\sigma}c_{i\sigma} \label{H1}
\end{eqnarray}

\noindent where $c^\dagger_{i\sigma} (c_{i\sigma})$ creates
(destroys) a Fermi atom with pseudospin $\sigma = \uparrow,
\downarrow$ at lattice site $i$. The density operator is $n_{i \sigma}=
c^\dagger_{i\sigma} c_{i\sigma}$ and $<ij>$ indicates the nearest
neighbor pair of sites. The average chemical potential $\mu =(\mu_\uparrow +\mu_\downarrow)/2$ and the chemical potential difference $h = (\mu_\uparrow - \mu_\downarrow)/2$, where $\mu_\sigma$ is the chemical potential of hyperfine state $\sigma$. The on-site interaction $U$ can be repulsive or attractive. Its magnitude and sign can be controlled by the \emph{s}-wave scattering length of the two Fermi species. In the present work, we consider only the positive $U$. In cold-gas experiments, the ratio $U/t$
is controlled by the intensity of the standing laser waves (the tunneling amplitude $t$
is exponentially sensitive to the laser intensity while the on-site interaction $U$ is weakly sensitive). Experimentally, the 1D geometry can be realized by a strong confinement in the transverse direction with an additional periodic potential applied along
the other direction. We consider a tight one dimensional geometry such that the level spacing in transverse direction is much
larger than the energy per particle of the axial direction.

In the presence of the underlying harmonic oscillator potential, the average chemical potential $\mu(z)$ monotonically decreases from the center to the edge of the trap. The spin diffusion current driven by this spatial variation of the chemical potential is given by the modified Fick's law~\cite{spte9};

\begin{eqnarray}
j_m = -D \chi \frac{\partial (M/\chi)}{\partial z}\label{FL}
\end{eqnarray}

\noindent where the magnetization density $M(z) = n_\uparrow(z) - n_\downarrow(z)$, the spin susceptibility $\chi(z) = \partial M/\partial h$, and the spin diffusion coefficient $D(z)$ are values that vary in space, even at equilibrium. Here $n_\sigma(z)$ is the density of hyperfine state $\sigma$ at position $z$. First, we solve the homogeneous Hubbard model for the population imbalance system using the thermodynamic bethe ansatz (TBA) technique
and numerically calculate the magnetization and the spin susceptibility as discussed in Refs.~\cite{andrew, tak}. By combining the TBA solutions with the local density approximation (LDA), we then extract the local quantities $m(z)$ and  $\chi(z)$. In LDA, the external trapping potential, which is independent of hyperfine state, is $V_{i} =
m\omega^2 z^2/2$ at site $i$. This is related to the local chemical potential through the relation $\mu_i = \mu_0 - V_i$, where $\mu_0$ is the central chemical potential and $z = id$, with lattice constant $d$ is the spatial coordinate. We notice that while $\mu_{z}$ is spatially dependent due to LDA, h is not and remains constant.

\section{III. The results}

The spin current given in Eq. (\ref{FL}) is driven by the density imbalance and is a longer timescale result. There are several time scales associated with the experimental setup. The equilibrium time scale is the maximum of $\hbar/U$ or $\hbar/t$. There are two other larger time scales, one associated with trapping frequency $\pi/\omega$ and the other associated with the lattice $\lambda/(2 v_F)$, where $v_F$ is the fermi velocity. The time scale we consider here is $T > \pi/\omega \gg \hbar/t$ . Systems of trapped fermi atoms without the presence of an optical lattice have been studied experimentally in Refs.~\cite{spte3, spte6}, where $T > \pi/\omega$, thus, the addition of a lattice with the timescale parameter regime being $T > \pi/\omega \sim \lambda/(2 v_F) \gg \hbar/t$ is feasible in current setups.

\subsection{The spin current}

Using LDA, the derivative with respect to spatial coordinate can be converted into a derivative over the average chemical potential $\mu$. Noticing that $h$ is independent of the spatial coordinate and using the relations $\partial/\partial z = (\partial \mu/\partial z)\partial/\partial \mu$ and $(\partial \mu/\partial z) = -m\omega^2 z$, spatial dependence of the ratio of spin current and diffusion coefficient is written as

\begin{eqnarray}
j_m/D = m\omega^2 z\biggr[\chi \frac{\partial M}{\partial \mu}-M \frac{\partial \chi}{\partial \mu}\biggr].
\end{eqnarray}

\begin{figure*}
\includegraphics[width=\textwidth,clip]{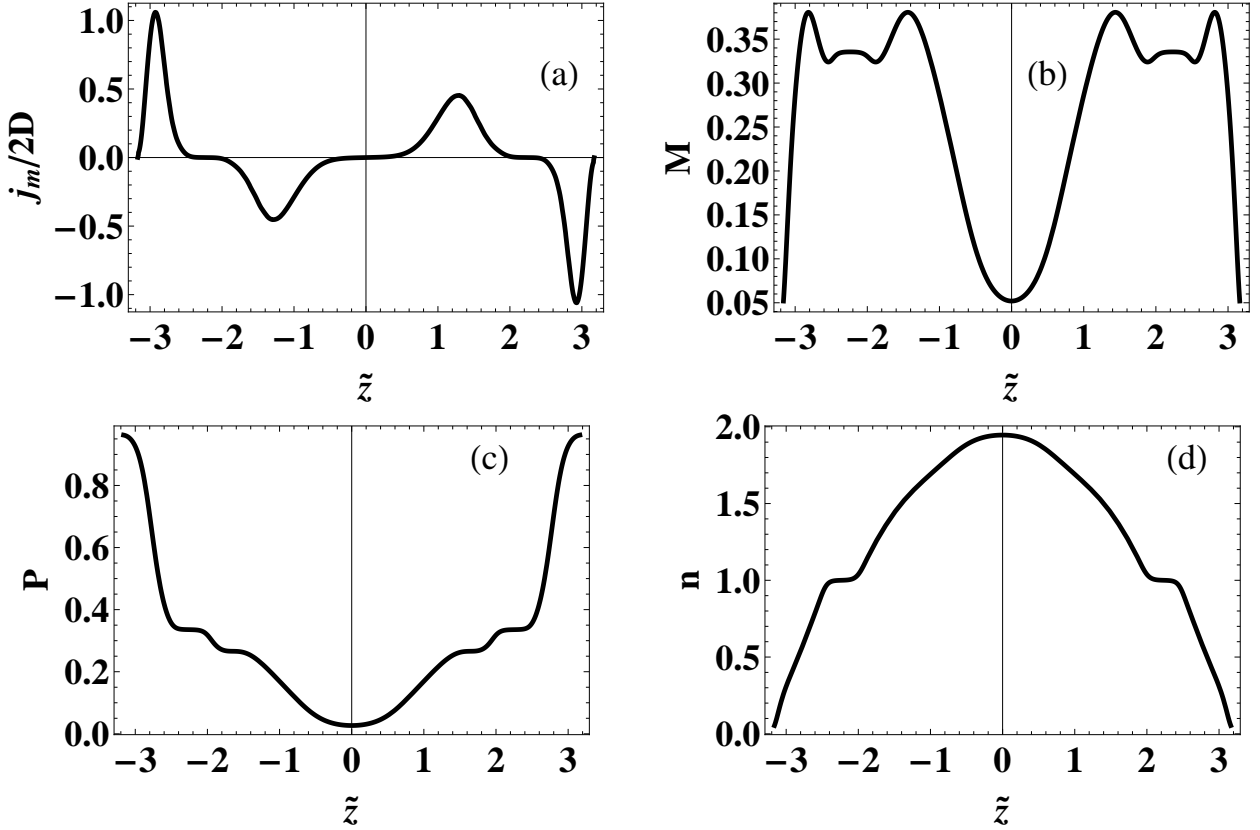}
\caption{For strong coupling, the spatial variations of $j_m(z)/D(z)$ (where $j_m(z)$ is the spin current and $D(z)$ is the spin diffusion coefficient), magnetization $M(z) = n_\uparrow(z) - n_\downarrow(z)$, polarization $P(z) = [n_\uparrow(z) - n_\downarrow(z)]/[n_\uparrow(z) + n_\downarrow(z)]$, and atom density $n(z) = n_\uparrow(z) + n_\downarrow(z)$ are shown in panel (a)-(d) respectively. We define the scaled length $\tilde{z} = z\sqrt{m\omega^2/2}$. Quantities plotted are dimensionless. We fixed the on-site interaction ($U =5t$), the inverse temperature ($\beta = 5/t$), and the chemical potential difference ($h = 0.4t$).}\label{spc}
\end{figure*}

\noindent Figure 1 shows the results for relatively low temperature and large on-site interactions where Mott insulating state is present in the middle of the trap. We find that all the local quantities, including spin current, show a considerable spatial variation even in equilibrium (see FIG. \ref{spc} caption for the details).

\begin{figure}
\includegraphics[width=\columnwidth]{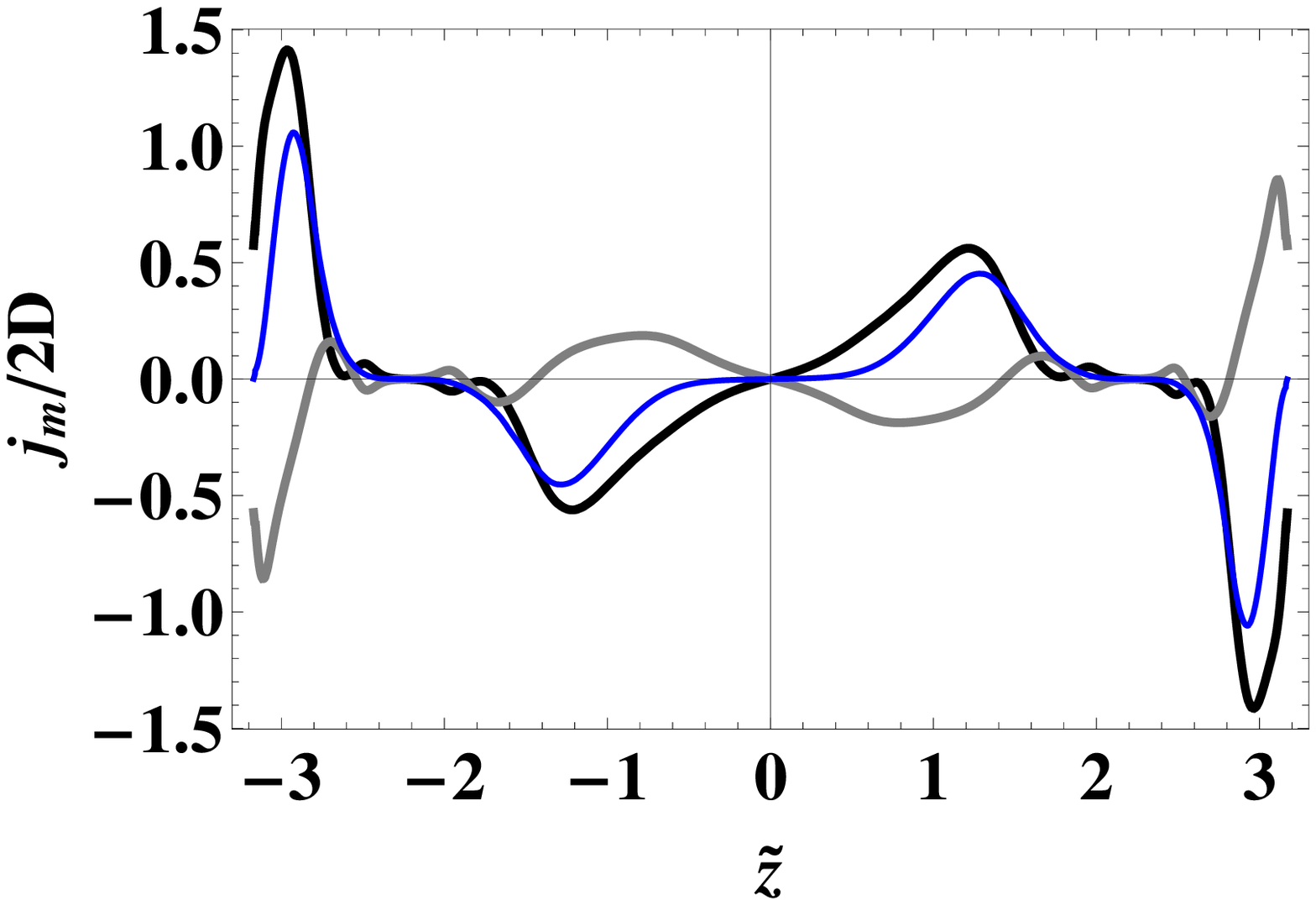}
\caption{[color online] The dimensionless spin current for the same parameters as in FIG.~\ref{spc}. The total spin current (blue dashed line) $j_m = j_1 +j_2$ are sum of two contributions. The magnetic contribution (gray line) is $j_1/2D = -\partial M/\partial z$ and the susceptibility contribution (black line) is $j_1/2D = (M/\chi) \partial \chi/\partial z$}\label{com}
\end{figure}

As seen in FIG. \ref{spc}, the magnetization is smallest at the edge and the center of the trap. The spin current flows from low magnetization to high magnetization regions. This current pattern originates from two contributions as shown in FIG.~\ref{com} and the atoms diffuse towards the Mott insulating region. On the other hand, the local diffusion arises from the collisions with opposite spin atoms. As the time between scattering event $\tau$ is proportional to the local spin diffusion coefficient ($D$), $D$ becomes very large when the density $n$ and holes $(1-n)$ are small. This is the reason why the ratio of spin current to diffusion coefficient $(j_m/D)$ is almost zero at the center and at the edges. The magnetization is almost constant in the Mott insulator region, so that $(j_m/D)$ is zero in this region. A similar current pattern (from edge towards the center) has been observed for an attractively interacting one dimensional non-lattice fermions~\cite{spte4, spte6}.

\begin{figure*}
\includegraphics[width=\textwidth,clip]{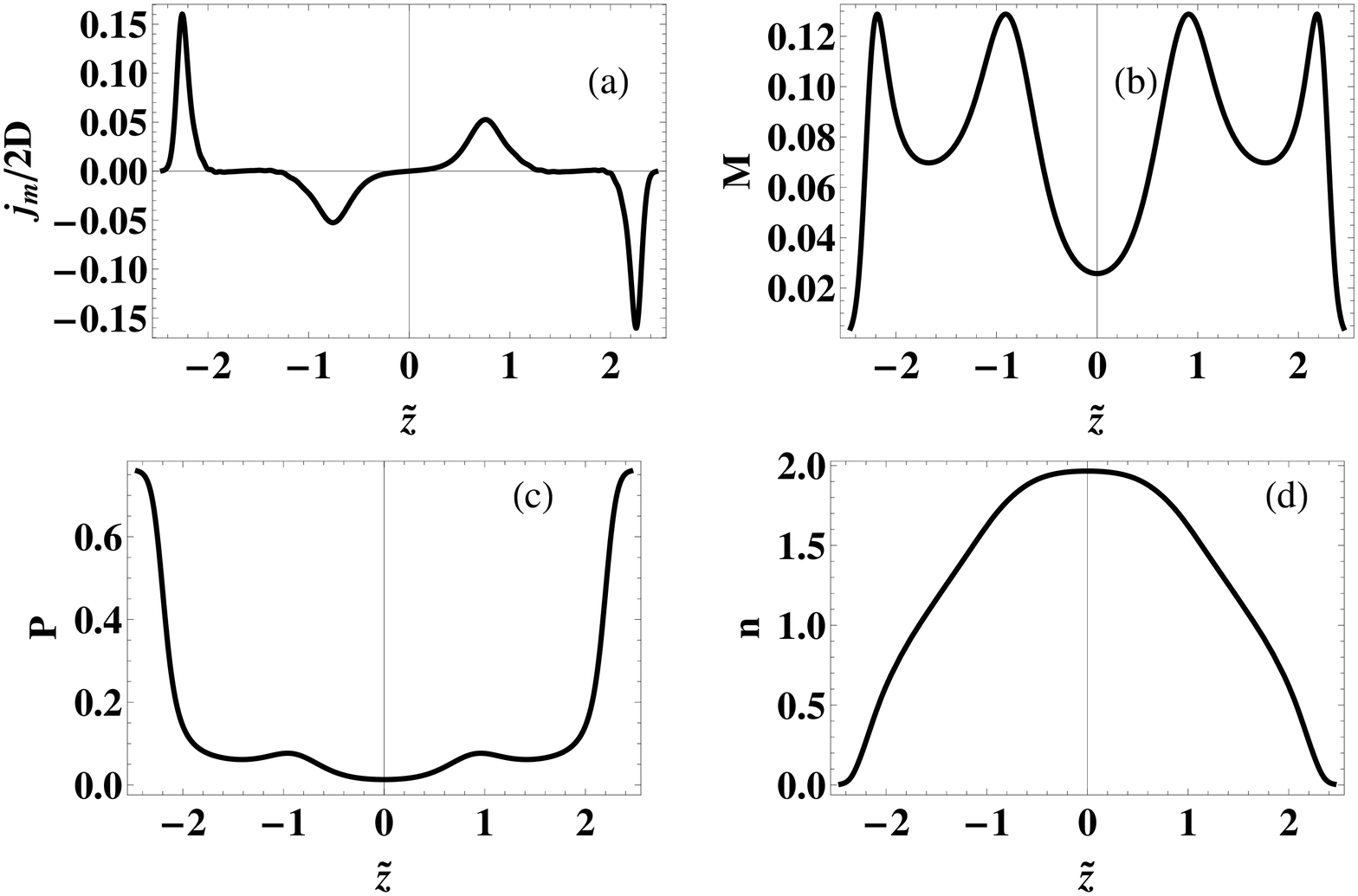}
\caption{For weak coupling, the spatial variations of $j_m(z)/D(z)$ (where $j_m(z)$ is the spin current and $D(z)$ is the spin diffusion coefficient), magnetization $M(z) = n_\uparrow(z) - n_\downarrow(z)$, polarization $P(z) = [n_\uparrow(z) - n_\downarrow(z)]/[n_\uparrow(z) + n_\downarrow(z)]$, and atom density $n(z) = n_\uparrow(z) + n_\downarrow(z)$ are shown in panel (a)-(d) respectively. We define the scaled length $\tilde{z} = z\sqrt{m\omega^2/2}$. Quantities plotted are dimensionless. We fixed the on-site interaction ($U =0.8t$), the inverse temperature ($\beta = 5/t$), and the chemical potential difference ($h = 0.2t$).  }\label{spc2}
\end{figure*}

The results for a weak coupling limit is shown in FIG.~\ref{spc2}. We find few qualitative differences other than the disappearance of Mott insulating region. As one expects, Mott region disappears at high temperatures and small interactions but the current pattern remains the same. We find that the spatial variations of these quantities are greater for larger population imbalance (or larger values of $h$).

\subsection{The local spin diffusion coefficient}

The fluctuation-dissipation theorem

\begin{eqnarray}
\chi^{\prime \prime}(k, \omega) = \frac{1}{2\hbar}(1 - e^{\beta\hbar \omega}) S(k, \omega) \label{fdt}
\end{eqnarray}

\noindent where the inverse temperature $\beta = \/(k_BT)^{-1}$, and the imaginary part of the dynamic magnetization response function

\begin{eqnarray}
\chi^{\prime \prime}(k, \omega) = \frac{1}{2}\int_{-\infty}^{\infty} dt \int dze^{i\omega t-i kz} \langle [\hat{M}(z,t),\hat{M}(0,0)]\rangle,
\end{eqnarray}

\noindent is related to the magnetization correlation function

\begin{eqnarray}
S(k, \omega) = \sum_i\int dt e^{i\omega t- i kz} S(z,t).
\end{eqnarray}

\noindent Here $S(z,t) = \langle \hat{M}(z,t)\hat{M}(0,0)\rangle$ and $[A,B]$ represents the commutator between the operators $A$ and $B$.
$ \langle \hat{M} \rangle = M$ represents the expectation value with respect to the Hamiltonian $H$.

First, we use general hydrodynamics description to derive the magnetization~\cite{book1}. Inserting the spin current $j_m$ into the continuity equation $\partial_t M+ \partial_z j_m = 0$, the diffusion equation has the form $\partial_t M(z,t) = D \chi \nabla^2 [M(z,t)/\chi]$. Multiplying this by $e^{i\omega t}$ and integrating the left hand side, and taking Laplace transformation and Fourier transformation, the solution of the diffusion equation can be written as
\begin{eqnarray}
M(k, \omega) = \frac{i}{\omega + iDk^2}M(k, t=0)\label{solDE}.
\end{eqnarray}

\noindent Notice that the diffusion process is reflected by the diffusion pole on the negative imaginary axis at $\omega = -iDk^2$. As the diffusion coefficient is local, the diffusion pole may not be as simple as it looks.

In order to establish the connection between the hydrodynamic diffusion equation and the dynamic magnetization response function, we assume that spatially varying chemical potential difference $h(z, t) = h(z) e^{\epsilon t}$ is adiabatically turned on to mechanically induce a non-zero magnetization. At time $t =0$, the chemical potential difference is switched off and allow the induced magnetization to be relaxed as the system returns to equilibrium. The induced magnetization at $t = 0$ is

\begin{eqnarray}
M(z, t=0) =2i\int_0^\infty d\tau \int dz^{\prime} \chi^{\prime \prime}(z-z^\prime, \tau) e^{-\epsilon \tau}h(z^\prime).
\end{eqnarray}

\noindent The Fourier transform of this equation has the form $M(k, t=0) = \chi(k) h(k)$, where

\begin{eqnarray}
\chi(k) = \int \frac{d\omega}{\pi \omega}\chi^{\prime \prime}(k, \omega).
\end{eqnarray}

\noindent For $t > 0$, the Laplace transformation of the induced magnetization
\begin{eqnarray}
M(z, t) =2i\int_{-\infty}^0 d\tau \int dz^\prime \chi^{\prime \prime}(z-z^\prime, t-\tau) e^{\epsilon \tau}h(z^\prime),
\end{eqnarray}

\noindent has the form
\begin{eqnarray}
M(k, \eta) = \int \frac{d\omega}{\pi i} \frac{\chi^{\prime \prime}(k,\omega)}{\omega (\omega-\eta)}h(k).
\end{eqnarray}

\noindent Using $h(k) = M(k, t=0)/\chi(k)$ and

\begin{eqnarray}
\chi(k,\eta) = \int \frac{d\omega}{\pi} \frac{\chi^{\prime \prime}(k,\omega)}{(\omega-\eta)},
\end{eqnarray}

\noindent we find
\begin{eqnarray}
M(k, \eta) = \frac{1}{i\eta} \biggr[\frac{\chi(k,\eta)}{\chi(k)} -1 \biggr]M(k, t=0).
\end{eqnarray}

\noindent Comparing this with Eq. (\ref{solDE}), the dynamical magnetization response function can be extracted

\begin{eqnarray}
\chi(k,\eta) = \frac{iDk^2}{\eta + iDk^2}\chi(k).
\end{eqnarray}

\noindent Setting $\eta = \omega +i \epsilon$ and taking the imaginary part at the limit of $\epsilon \rightarrow 0$, we get
\begin{eqnarray}
\chi^{\prime \prime}(k,\omega) = \frac{Dk^2\omega}{\omega^2+(Dk^2)^2}\chi(k)\label{imchi}.
\end{eqnarray}

\noindent Since $\chi^{\prime \prime}(k,\omega)$ is related to the magnetization correlation function $S(k,\omega)$ through the Eq. (\ref{fdt}), we notice that $S(k,\omega)$ gives a quasielastic peak with width $\Gamma(k) = 2Dk^2$.

On the other hand, the spin conductivity $\tilde{D}(\omega)$ can be written in terms of the spin current $I_s(k,t)$

\begin{eqnarray}
\tilde{D}(\omega) &=& \frac{2\omega}{\chi} \lim_{k\to 0} \frac{1}{k^2}\chi^{\prime \prime}(k,\omega) \\ \nonumber
           &=& \frac{2a^2}{\omega \chi} Im \Pi(\omega)\label{scond}.
\end{eqnarray}

\noindent Here the spin current response function is
\begin{eqnarray}
\Pi(\omega) = -\frac{i}{N}\lim_{k\to 0} \int_0^\infty dt e^{i\omega t} \langle [I_s(k,t),I_s(-k,0)]\rangle,
\end{eqnarray}

\noindent where the spin current operator $I(k, t) =\sum_{k^\prime,\sigma}\sigma\sin (k^\prime d)c^\dagger_{k^\prime,\sigma}c_{k^\prime +k,\sigma}$. Combining Eqs. (\ref{imchi}) and (\ref{scond}), the spin diffusion coefficient can be related to the spin conductivity as $D = \tilde{D}(0)/2$~\cite{fishman}. Following the reference~\cite{peter}, the \emph{long-time} spin diffusion coefficient of the one dimensional Hubbard model can be casted into

\begin{eqnarray}
D = \frac{d^2t^2}{\hbar \chi k_B T}\frac{C_0}{2}\biggr[\frac{2 \pi C_0}{C_2}\biggr]^{\frac{1}{2}},
\end{eqnarray}

\begin{figure}
\includegraphics[width=\columnwidth]{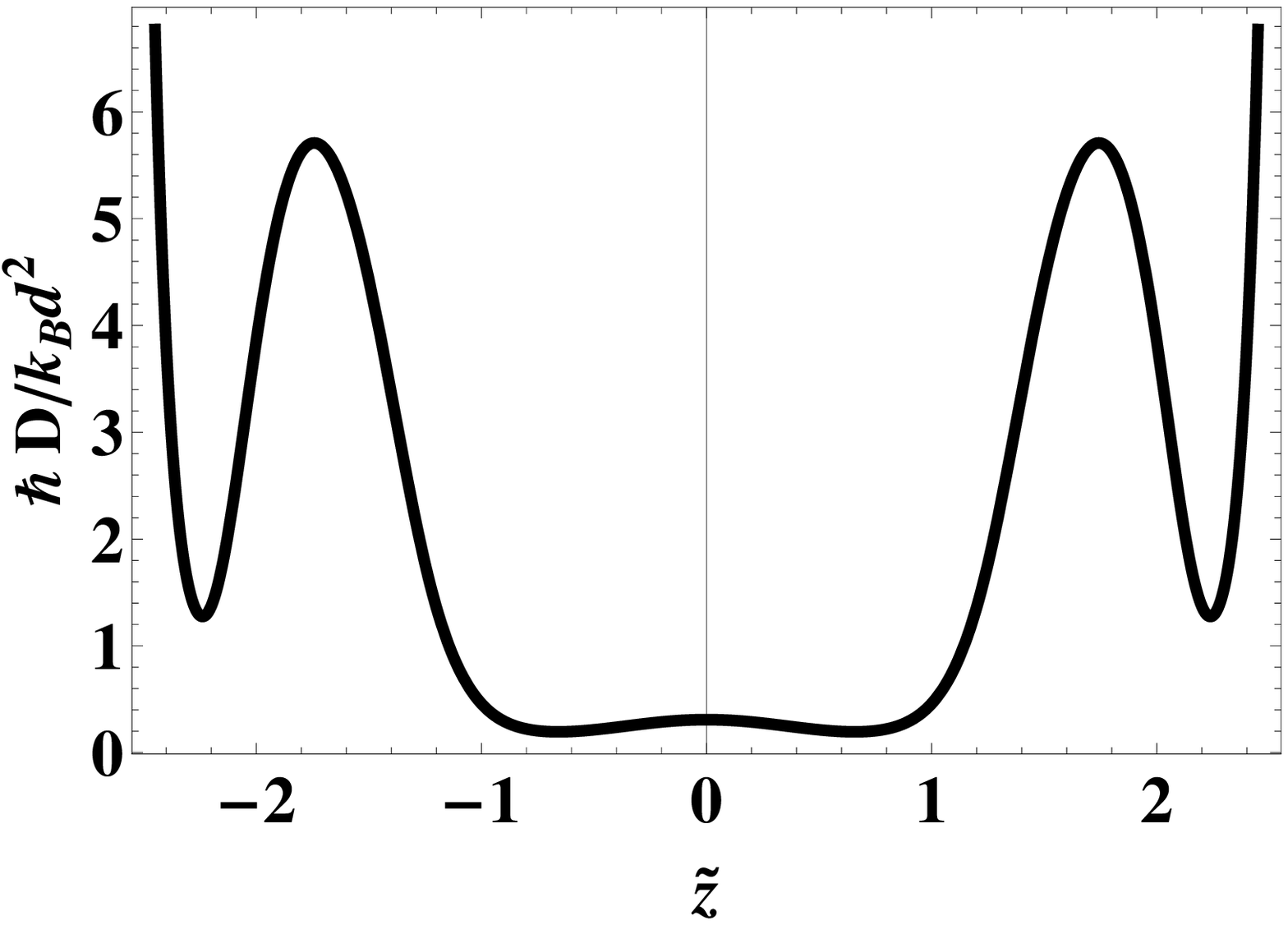}
\caption{Diffusion coefficient as a function of position for the weak coupling case. The scaled length $\tilde{z} = z\sqrt{m\omega^2/2}$, the on-site interaction ($U =0.8t$), the inverse temperature ($\beta = 5/t$), and the chemical potential difference ($h = 0.2t$). }\label{dvsz}
\end{figure}

\noindent where $d$ is the lattice constant and the coefficients $C_0 = \langle I^2(0,0) \rangle$ and $C_2 = \langle I(0,0)[[I(0,0),H],H] \rangle$ are the first two \emph{short-time} expansion coefficients of the current correlation function. Note that the \emph{long-time} spin diffusion coefficient is written in terms of two lowest order \emph{short-time} expansion coefficients of the current correlation function ~\cite{peter}.

In the \emph{weak coupling} limit, by generalizing the results in Ref.~\cite{peter}, we find

\begin{eqnarray}
 C_0 =\biggr[\frac{1}{N}\sum_k \sin (kd) (n_{k\uparrow} - n_{k\downarrow})\biggr]^2 \\ \nonumber
+ \frac{1}{N}\sum_{k\sigma} \sin^2 (kd)n_{k\sigma}(1-n_{k\sigma})
\end{eqnarray}

\noindent and
\begin{eqnarray}
C_2 = U^2C_0\biggr\{\biggr[\frac{1}{N}\sum_k (n_{k\uparrow} - n_{k\downarrow})\biggr]^2 \\ \nonumber
+ \frac{1}{N}\sum_{k\sigma} n_{k\sigma}(1-n_{k\sigma})\biggr\}
\end{eqnarray}

\noindent Here the Fermi function $n_{k\sigma} = (e^{\beta\epsilon_{k\sigma}}+1)^{-1}$, and $\epsilon_{k\sigma} = 2t\cos(kd)-\mu_\sigma$. The spatial dependance of the diffusion coefficient enters through the local chemical potential $\mu_{i\sigma} = \mu_{0\sigma} -\gamma i^2$.

As shown in Fig. 4, we find the diffusion coefficient to be small through the middle of the trap, though there is a local maximum at the center of the trap as predicted since the trap center is occupied by both up and down spin particles. A maximum is also seen when the local density is equal to unity, a remnant of the Mott transition, as well as at the edge of the trap, where the density is almost zero, showing that D is large when density and holes are small. Further, the spin diffusion coefficient tends to reach a local maximum when the magnetization is a local minimum, and visa versa. By combining FIG. 3(a) and FIG. 4, one can see that the spin current is large at the edges of the trap in the lower metallic region, finite but small in the upper metallic region, and zero in the insulating regions.

\subsection{ The Damping rate}

Taking the inverse Laplace transformation of Eq. (\ref{solDE}), one can write
\begin{eqnarray}
M(k, t) = e^{-Dk^2t}M(k, t=0)\label{dpr}.
\end{eqnarray}

\noindent This shows that the damping rate of hydrodynamic diffusion mode $\tau(k) = 1/(Dk^2)$ is consistent with the quasielastic peak of $S(k, \omega)$. As the diffusion coefficient $D = D(k)$ for trapped fermions, Eq. (\ref{dpr}) does not give us much information on the damping rate of the diffusion mode. However, the damping rate of the diffusion modes can be estimated directly from the continuity equation by assuming $M(z, t)$ has the form $M(z, t) = \exp[-\tau t]M(z)$. Inserting this into the continuity equation and integrating both sides, we find

\begin{eqnarray}
\tau = \frac{j_m(z_f)-j_m(z_i)}{\int_{z_i}^{z_f} M(z) dz}.
\end{eqnarray}

\noindent When estimating $\tau$ using this equation, one must consider only a part of the metallic cloud with edges $\tilde{z}_i$ and $\tilde{z}_j$, since the spin does not transport through an insulating region. We estimate $\tau$ from a small region in the lower metallic band, from  $\tilde{z}_i=2.21$ to $\tilde{z}_j=2.42$, and find $\tau \approx 21 \hbar/(k_Bd^2)$.

\section{IV. Discussion and Summary}

The local spin current patterns predicted above for neutral atomic fermions in an optical lattice can experimentally be probed by measuring lattice scale modulations of the atom density. Noise correlations~\cite{nc1, nc2}, Bragg scattering~\cite{bs}, and \emph{in situ} imaging in the lattice scaling~\cite{insm} are commonly used density mapping tools in cold atom experiments. While Bragg scattering is similar to X-ray scattering probes of the crystal structures of solids, \emph{in situ} imaging is analogous to scanning tunneling microscopy. Imaging the subsequent lattice scale density modulations as done for one dimensional Bose gases~\cite{1DBG}, local spin current patterns can be probed.

In this paper we have found evidence of longer timescale spin current in both the strong and weak coupling cases. By calculating the ratio $j_m(z)/D(z)$, we find that the spin current is driven primarily by susceptibility rather than directly from magnetization density. In the weak coupling limit, we further found how the local spin diffusion coefficient varies within the trap, having maximums at the insulating regions. We find spin current to be primarily located edges of the trap. Finally, we estimated the damping rate of diffusion modes in a metallic portion of the cloud.

\section{V. Acknowledgements}

AS would like to thank Patrick O'Brien for critical discussions and assistance with Mathematica.

\end{document}